\documentclass[graybox, nosecnum]{svmult}

\usepackage{mathptmx}       % selects Times Roman as basic font
\usepackage{helvet}         % selects Helvetica as sans-serif font
\usepackage{courier}        % selects Courier as typewriter font
\usepackage{type1cm}        % activate if the above 3 fonts are
\usepackage{tabularx}
\usepackage{array}
\usepackage{makeidx}         % allows index generation
\usepackage{graphicx}        % standard LaTeX graphics tool
\usepackage{multicol}        % used for the two-column index
\usepackage[bottom]{footmisc}% places footnotes at page bottom
\usepackage{hyperref}        %for hyperlinks
\usepackage{soul}            % for high-lighting of text
\hypersetup{colorlinks=true,urlcolor=blue}
\usepackage[square,numbers]{natbib}
\makeindex             % used for the subject index
\begin{document}
%\tableofcontents{}
\title*{CubeSats for Gamma--Ray Astronomy}
\author{Peter Bloser \thanks{corresponding author}, David Murphy \thanks{corresponding author}, Fabrizio Fiore and Jeremy Perkins}
\authorrunning{P. Bloser et al.} %for an abbreviated version of
\institute{Peter Bloser \at Los Alamos National Laboratory, Los Alamos NM 87545, USA, \email{pbloser@lanl.gov}
\and David Murphy \at Centre for Space Research and School of Physics, University College Dublin, Ireland \email{david.murphy@ucd.ie}
\and Fabrizio Fiore \at INAF, Osservatorio Astronomico di Trieste, Italy \email{fabrizio.fiore@inaf.it}
\and Jeremy Perkins \at NASA/GSFC, Greenbelt MD 20771, USA  \email{jeremy.s.perkins@nasa.gov}
}
\maketitle
\abstract{After many years of flying in space primarily for educational purposes, CubeSats - tiny satellites with form factors corresponding to arrangements of ``1U'' units, or cubes, each 10 cm on a side - have come into their own as valuable platforms for technology advancement and scientific investigations.  CubeSats offer comparatively rapid, low-cost access to space for payloads that be built, tested, and operated by relatively small teams, with substantial contributions from students and early career researchers.  Continuing advances in compact, low-power detectors, readout electronics, and flight computers have now enabled X-ray and gamma-ray sensing payloads that can fit within the constraints of CubeSat missions, permitting in-orbit demonstrations of new techniques and innovative high-energy astronomy observations.  Gamma-ray-sensing CubeSats are certain to make an important contribution in the new era of multi-messenger, time-domain astronomy by detecting and localizing bright transients such as gamma--ray bursts, solar flares, and terrestrial gamma--ray flashes; however, other astrophysical science areas requiring long observations in a low-background environment, including gamma--ray polarimetry, studies of nuclear lines, and measurement of diffuse backgrounds, will likely benefit as well.  We present the primary benefits of CubeSats for high-energy astronomy, highlight the scientific areas currently or soon to be studied, and review the missions that are  currently operating, under development, or proposed.  A rich portfolio of CubeSats for gamma--ray astronomy already exists, and the potential for a broad range of creative and scientifically productive missions in the near future is very high.  }

% Please provide keywords required to facilitate search of chapter on web; maximum 10 keywords.
\section{Keywords} CubeSats; Gamma--ray Bursts; Solar Flares; Terrestrial Gamma--ray Flashes; Multi-messenger astronomy; Transients; Nanosatellites; Constellations

\newpage

\section{Introduction}
%- What are nanosats and cubesats?

A nanosatellite is normally considered a spacecraft of mass less than 10\,kg. A CubeSat is a nanosatellite built from cubes (1U), 10\,cm on a side, which approximately conforms to the Cal Poly CubeSat specification that was originally developed for educational reasons\footnote{\url{www.cubesat.org/cubesatinfo}}. This specification allocates a mass of $\sim1.33$\,kg per 1U volume for most CubeSat sizes, although the 6U and 12U specifications allow for total masses of 12 kg and 25 kg, respectively. More than 2,000 nanosatellites have been launched to date, with 3U CubeSats being the most common\footnote{\url{www.nanosats.eu}}. CubeSat launch opportunities are available through rideshares at a fraction of the cost of a traditional single payload launch. Tens, or even hundreds, of CubeSats may be released from a given launch vehicle, typically from a dispenser or 'deployer', but sometimes manually by astronauts from the International Space Station (ISS). A new wave of small-lift launch vehicle manufacturers, such as Relativity, Rocket Lab and Astra, is catering to the growing demand for small satellite launch opportunities \cite{Kulu2021}. 

Nanosatellites and CubeSats can be used as platforms for in-orbit demonstration (IOD), accelerating the qualification cycle of new technologies.  Commercial off-the-shelf technology originally developed for non-space sectors can also be rapidly trialled for space applications using CubeSats. For example, the Intel Movidius Myriad 2 AI/Edge processor has been successfully used on the Phisat-1 Earth Observation mission for on-board cloud detection \cite{Giuffrida2020}.

In parallel to their usefulness for IOD, the capabilities of nanosatellites for many areas of scientific research are of rapidly growing interest. CubeSats are flexible platforms that can be configured for a wide range of science mission profiles, either as a standalone platform, as a communications relay for lunar and inter-planetary missions, or as a daughter spacecraft to study a near-Earth object \cite{cubesathandbook2021}. One example is the 6U LICIACube satellite that piggy-backed on NASA's DART mission to record images of the impact of the main spacecraft with the asteroid Dimorphos \cite{dotto2021}. 

Although the primary use of CubeSats in swarms or constellations so far has been primarily for Earth Observation and telecommunications, such a use case can be foreseen for other application areas. CubeSat constellations will likely grow in number as they can potentially improve our understanding of the space environment with their ability to capture simultaneous, multi-point measurements with identical instruments over a large area \cite{Burkhard2021}. A COSPAR scientific roadmap found that all branches of space science can potentially benefit from the smaller envelope, lower cost, and higher repeat rate achievable with small satellites, with particular benefit to scientific communities in small countries \cite{Millan2019}. 

Educational programmes involving CubeSats provide valuable hands-on, full life-cycle space mission experience to university and high school students \cite{Castillo2021,Kinnaird2022,Schloms2022}. More than 390 students so far have worked on the GRID (n=150) \cite{grid_wen}, IGOSat (n=200) \cite{Phan2018}, and EIRSAT-1 (n=40) \cite{murphy2021,murphy2022} gamma--ray CubeSat missions. 

Modern space-based gamma--ray instruments are typically massive (several hundred kg) and highly complex. The penetrating power of gamma--rays requires the use of relatively thick, high-Z, materials to maximise the interaction probability. The high background levels in orbit usually require shielding of the electronics. The lack of focusing capability means that the detector area is comparable to the collecting area.  One of the major technical challenges to accessing the relatively unexplored energy band from 100\,keV to several GeV with the required sensitivity is the need to exploit both Compton scattering and pair creation in a single instrument, which leads to significant mass and additional complexity in both hardware and on-board software. \textit{ Nevertheless, due to recent advances in compact detector and readout electronics technologies, CubeSats have begun to fill an important need in gamma--ray astronomy: rapid, low-cost access to space for both technology demonstration and targeted science investigations.}  CubeSats can play an important role by demonstrating the performance in-orbit of new technologies, advancing their Technology Readiness Level (TRL) to the point where major space agencies may consider their use in larger missions. The science case for CubeSats, especially in the Multi-Messenger Astronomy era, is also growing in breadth and impact, as described below.  

This chapter focuses on the main scientific topics in gamma--ray astronomy that can be addressed by CubeSats and presents examples of operational and planned CubeSat missions. An overview of CubeSat subsystems and payloads for gamma--ray detection is given in  \cite{Arneodo2021}.
Two successful CubeSat missions for X--ray astronomy, HaloSat and PolarLight, are described elsewhere in this Handbook \cite{FengKaaret2022}. A discussion of some enabling technologies for gamma--ray CubeSats, especially silicon photomultipliers (SiPMs), is presented in \cite{Bissaldi2022}. 

\section{CubeSats as Platforms for In-Orbit Demonstration (IOD) of New Technologies}
\label{sec:iod}
Despite numerous proposed missions over the past decade \cite{Lebrun2014,Greiner2012,vonBallmoos2012,Tatischeff2016,deAngelis2021,Kierans2020a}, development of a successor to the COMPTEL instrument on the Compton gamma--ray Observatory in the 1990's \cite{Schonfelder1993} has been hampered in part by the low TRL of the necessary instrumentation. Fast, high light-yield scintillators, compact optical detectors, such as SiPMs, and high-speed digital electronics, can all greatly improve instrument performance while reducing mass, volume, and complexity \cite{bloser2014}. 

Developing the necessary technologies, assessing their performance, and demonstrating their flight-worthiness in a representative configuration, are important steps for a successful technical feasibility assessment of any proposed mission concept to a space agency. High altitude balloons have long been used as test-beds for new space technologies and are part of the ``roadmap'' for new space mission development, especially at gamma--ray energies. The performance of a compact Compton telescope design, COSI, based on Germanium \cite{Beechert2022} (and approved for flight as a Small Explorer mission by NASA in 2025) has previously been demonstrated on a balloon flight, as has a Compton telescope based on SiPMs, with LaBr$_3$ \cite{Bloser2016} and CeBr$_3$ \cite{Sharma2020}. 

The United States Department of Defense Space Test Program (STP) has provided IOD of the compact SIRI instrument, combining a SiPM array with an SrI$_2$ scintillator and successors SIRI-2, GARI (using GAGG scintillator) and Glowbug are in development \cite{Mitchell2019,Mitchell2021}. The TRL of these new technologies has been advanced further with the launch of the GECAM pair of micro-satellites in December 2020 which use a hemispherical array of detectors composed of LaBr$_3$ with SiPMs, to detect gamma--ray bursts (GRBs) and other transients \cite{Chen2021}. Many of the CubeSats described in this paper have dual purpose as `pathfinders' for future science missions or constellations. For example, GRBAlpha and VZLUSAT-2 are pathfinders for the CAMELOT constellation \cite{ripa2022}.

Particularly important has been the assessment of SiPM radiation damage in-orbit, and the impact on performance. Laboratory studies have now been supplemented by a wealth of in-orbit measurements which are proving important in assessment of the suitability of SiPMs for use in larger missions with longer lifetimes  \cite{Bissaldi2022}.

\section{The Science Case for High-Energy Astrophysics CubeSats}
Table~\ref{table:survey} lists some of the planned and operational CubeSat missions of relevance for high-energy astrophysics. The main science goals are typically the study of bright transient events such as GRBs, solar flares and terrestrial gamma--ray flashes (TGFs). However, additional science goals that can benefit from long, uninterrupted observations, and from the relatively low instrumental background generated by energetic particles in a low-mass CubeSat platform, are also emerging. (We note that many of these missions also have educational and technology demonstration goals.)   In this Section we summarize the primary science goals being pursued by current and proposed CubeSat missions in high-energy astronomy, in addition to related science utilizing gamma--ray sensing technology.

{\footnotesize 
\begin{table}[h]
    \begin{tabular} { | l | c | l | c | l | }
    \hline
   Mission & Type & Objective(s) & Energy Range & Status  \\
   \hline
  % ALBATROS & Not a nanosat    &              &              &        &             \\
   BlackCAT \cite{Falcone2022}  &   6U  &   High-z GRBs; Transients  &  0.5--20\,keV   & Launch 2024   \\
  BurstCUBE \cite{perkins2020} & 6U  & GRBs     &  50\,keV--1\,MeV       &  Launch 2023     \\
 %  CATCH & Not a nanosat >25kg    &              &              &        &             \\
   COMCUBE \cite{Tatischeff2022a,laviron2021} &  6U   &    GRBs; Polarization  &    50\,keV--few\,MeV  &    Study phase      \\
   COMPOL \cite{Yang2020, laurent2022} &  3U   &   Cyg X-1 polarization    & 160\,keV to few MeV    & Launch 2026     \\
  % Daksha &    Not a nanosat &              &              &        &             \\
 CUSP  \cite{Fabiani2022}   & 6U   & Solar Flares; Polarization & 20--100\,keV & Study phase  \\ 
   EIRSAT-1 \cite{murphy2021,murphy2022} &  2U  & GRBs; Ed; Tech Demo  &  30\,keV-–2\,MeV     & Launch 2023      \\
  % GALI \cite{gali} & Not enough info available to include here - has a section later   &       GRBS  &         TBC     &       \\
  % GECAM & not a nanosat    &              &              &           \\
   GIFTS \cite{Ulyanov2022} &  6U   &  GRBs    &  50--300\,keV    & In development       \\
  % GLOWBUG &  not a nanosat   &              &              &                 \\
   GRID \cite{grid_wen} &   6U  & GRBs &   $\sim 10$\,keV--2\,MeV            &      Launched 10/18; 11/20; 02/22    \\
   GRBAlpha \cite{pal2020} &  1U   &  GRBs; Solar Flares  &   70\,keV–-890\,keV    &  Launched 03/2021         \\
   GTM \cite{Chang2022} & 2$\times$ 1U payloads &   GRBs   &    50\,keV--2\,MeV  & Launch 2024                  \\
% HaloSat &  6U   & X-ray mention in Intro         &              &                  \\
   HERMES TP/SP \cite{fiore,evangelista} &  7$\times$\,1U payload & GRBs   &    5--500\,keV   &   Launch 2023     \\ 
   IGOSat \cite{Phan2018}  &  3U   & $\gamma$--rays LEO; Ed  &  20\,keV--2\,MeV   &  Qualification phase             \\    
   IMPRESS \cite{setterburg2022} &  3U   & Solar Flares  &  4--100\,keV  &   Launch 2023           \\ 
  LECX \cite{lecx} & 2U    & GRBs  &  30--200\,keV  &   Launch 2023           \\
   LIGHT-1 \cite{DiGiovanni2019,Almazrouei2021} & 3U  & TGFs; Education  & 20\,keV--3\,MeV   &  Deployed 03/2022        \\
 %  LUVS &   can't find info on this one  &              &              &                  \\
   MAMBO \cite{mambo_blos1,mambo_blos2} &  12U  &    Cosmic diffuse background    &    0.3--10\,MeV   & Launch 2023            \\
   MeVCube \cite{Lucchetta2022} &  6U   &   Transients; GRBs        &     200\,keV--4\,MeV         & Study phase        \\
 %  MoonBEAM & Not a nanosat    &              &              &        &             \\
 % PolarLight & 1U &    x-ray, mention in Intro          &              &         \\
 Min-XSS1 \cite{Mason2016} &  3U   &   Solar Flares; Ed  &   0.4--30\,keV  & 05/2016 to 05/2017  \\
 Min-XSS2 \cite{Mason2020} & 3U &   Solar Flares; Ed  &   0.4--30\,keV  & Launched 12/2018    \\
 SOCRATES \cite{Delange2016} &  3U  & GRBs; Tech demo; Ed   &  200\,keV--1.3\,MeV  &  Deployed 02/2020           \\
 %  SPIRIT hosting HERMES &     &              &              &           \\
  % Starburst &    not a nanosat &              &              &        &             \\
 %  QUVIK & Not a nanosat     &              &              &        &             \\
VZLUSAT-2 \cite{ripa2022} &  3U & GRBs; Solar Flares  & 30--890\,keV  & Launched 01/2022   \\
   3UTRANSAT \cite{atteia2022} &  3$\times$ 3U   & GRBs        &  15--200\,keV        &  Study phase  \\
    \hline
    \end{tabular}
    \caption{Planned and operational CubeSats (or CubeSat-sized payloads) for high-energy astrophysics. The main science goals are typically the detection of transients such as GRBs, TGFs and solar flares.}
    \label{table:survey}
    \end{table}
    }
%\end{center}
\subsection{GRBs and Multi-Messenger Astronomy}
\label{sec:grbmm}

GRBs are the most luminous electromagnetic explosions in the universe,  producing most of their electromagnetic energy in the form of a burst of gamma--ray photons, with energies from hundreds of keV to a few MeV, that lasts from seconds to minutes, with variability timescales as short as a few milliseconds. Their huge luminosities are generated in the most highly relativistic jets known in nature with bulk Lorentz factors above 100. These properties, combined with their cosmological distances, mean that GRBs are powerful probes of the universe back to the first population of stars that were formed, as test-beds for fundamental physics, and as laboratories for matter and radiation under extreme physical conditions that cannot be reproduced on Earth \cite{Piran2004}. 

GRBs arise primarily through two distinct formation channels \cite{Kumar2015}. The spectrally softer events (E$_{\rm peak}$ $\sim$ 150\,keV) \cite{Poolakkil2021} that last longer than about 2\,s are most likely associated with a progenitor that is a massive star in the process of collapse to form a black hole. Spectrally harder (E$_{\rm peak}$ $\sim$ 450\,keV) GRBs shorter than 2\,s are more likely to be associated with the coalescence of two compact stellar remnants, either a neutron star - neutron star, or a neutron star - black hole.  These `short’ GRBs have long been considered as excellent electromagnetic counterpart candidates to gravitational wave sources \cite{Berger2014} and therefore play an important role in time domain and multi-messenger astronomy \cite{MuraseBartos2019}.  Indeed, the short-lived gamma--ray transient, GRB\,170817A, was the first detected electromagnetic counterpart to a gravitational wave source whose progenitor was a binary neutron star merger  \cite{abbott2017}. 

Improving instrument performance, while reducing development time, cost and complexity, are key enablers to ensure the gamma--ray transient sky remains under surveillance during gravitational wave observing runs. Aswell as GRBs, gamma--ray transients such as soft gamma--ray repeaters and magnetar giant flares are also potential targets for CubeSat-based detectors. Gamma--ray sensors on different nanosatellites in distinct orbits can provide all-sky GRB detection and localisation capability, supporting observations by larger missions and optimising the chances for discovery of electromagnetic counterparts to GW sources during gravitational wave observatory operational phases. CubeSats are relatively low cost and have short launch timescales, making them ideal candidates to bridge potential gaps in coverage by the large gamma--ray missions. It is worth recalling that the detectors on the Vela 5 and 6 satellites that discovered GRBs consisted of arrays of small (10\,cm$^3$) CsI scintillators distributed around the spacecraft to ensure isotropic coverage \cite{klebesadel73}. Early GRB spectra were obtained from a 30\,cm$^3$ CsI(Tl) instrument on the IMP-6 spacecraft \cite{cline73}. GRBs are distinctive for their brightness and short duration that make them relatively easy to detect even with small sensors. An array of differently oriented detectors on a single spacecraft can provide rough localization (tens of square degrees), while detectors on spacecraft in different orbits can localize by triangulation. These are not new concepts, but implementation on CubeSat platforms can enable a greater diversity and frequency of missions, developed in a more agile and responsive way as new discoveries are made. 

\subsection{Solar Flares}
Solar flares are impulsive, unpredictable eruptions that release large amounts of energy at the solar surface as a result of magnetic reconnection in the solar corona, which leads to direct heating of local plasma and acceleration of particles (both electrons and ions) to high energies. Flares produce emission throughout the electromagnetic spectrum up to gamma--ray energies, generating very high fluxes which can lead to pulse pile-up effects, where two low-energy counts are incorrectly recorded as a single high-energy count. 

Debate is ongoing as to the mechanism of particle acceleration in solar flares, which is also important for understanding stellar processes and space weather \cite{Zharkova2011}. Observations of the bremsstrahlung hard X--rays (HXR) can reveal the location and energy spectra of the flare-accelerated electrons, while nuclear gamma--rays provide information on the accelerated protons and heavier ions \cite{Dennis2022}. 

Solar flares are highly variable in time, often displaying quasi-periodic behavior on short timescales.  One way to probe particle acceleration mechanisms is by investigating short-timescale (tens of milliseconds) variations in solar HXR observations, known as HXR spikes.  

\subsection{Terrestrial Gamma--Ray Flashes}
Terrestrial gamma--ray flashes (TGFs) are bursts of gamma--rays generated by electrons accelerated to relativistic energies in electric fields associated with lightning in Earth's atmosphere and were discovered by BATSE on CGRO in 1994 \cite{Fishman1994}. The median TGF duration is less than 100\,$\mu$s, and TGFs as short as 20\,$\mu$s have been reported. TGFs consist of photons with individual energies ranging from $< 10$ keV to $> 40$ MeV.  

Being very bright and easily observable from low Earth orbit (LEO), TGFs are a natural target for a small instrument suitable for CubeSat deployment.  The ASIM X- and gamma--ray instruments on the Columbus module of the ISS have a relative timing accuracy of 10\,$\mu$s and cover the energy range from 15\,keV to 20\,MeV \cite{Neubert2019}. TARANIS was a microsatellite designed for TGF studies that failed to reach orbit in 2020 \cite{Laurent2019}. 

\subsection{Persistent Sources}

CubeSats are well-suited to long, uninterrupted observations of persistent cosmic sources, a concept of operations that would be uneconomical for large, expensive missions with multiple science objectives.  Combined with advanced detectors and the inherently small instrumental background achievable on a low-mass CubeSat, this enables the high signal-to-noise measurements needed for studies of gamma--ray polarimetry, nuclear lines, and diffuse backgrounds.  CubeSat missions can thus play a valuable role doing path-finding science for relatively bright sources in these areas, and can do so long before the next generation of large gamma-ray missions is launched.

\subsubsection{Instrumental Background}

The sensitivity of any space-based gamma--ray sensor is limited by locally generated instrumental backgrounds, which are copiously produced by the interaction of energetic particles with spacecraft and instrument materials.  This limitation is especially severe for instruments operating near the $\sim0.5-10$\,MeV energy band, as shielding becomes difficult and heavy at these energies, and excited nuclei readily emit gamma rays in this range. It is also important for instruments without an anticoincidence shield, e.g. for the growing number of CubeSats which are being proposed for scientific missions. Furthermore, CubeSats for gamma--ray astronomy are often placed in non-optimal radiation environments as they are placed on ``low-cost'' rideshares whose main passengers are often Earth Observation missions.  (A comprehensive description of the background radiation encountered by gamma--ray instruments in space is given elsewhere in this Handbook \cite{Tatischeff2022b}.)

Because most discrete astrophysical gamma--ray sources are faint, the traditional approach to MeV astronomy relies on placing instruments with large collecting areas into space.  This results in massive (10$^3$--10$^4$\,kg) spacecraft busses and instrument structures; NASA’s Compton Gamma--ray Observatory (CGRO) had a mass of 17,000 kg.  Energetic particles in space, including galactic cosmic rays and solar energetic particles, interact with these massive structures (aluminum in particular) to produce an intense background “fog” of secondary gamma rays and neutrons. This instrumental background can be prompt or delayed (i.e. resulting from activation), and it severely impacts the sensitivity of space-based instruments (e.g. \cite{weidens_comptel_back},  \cite{weidens_spi_back}).

\begin{figure}[t]
\centering
    \includegraphics[width=\textwidth]{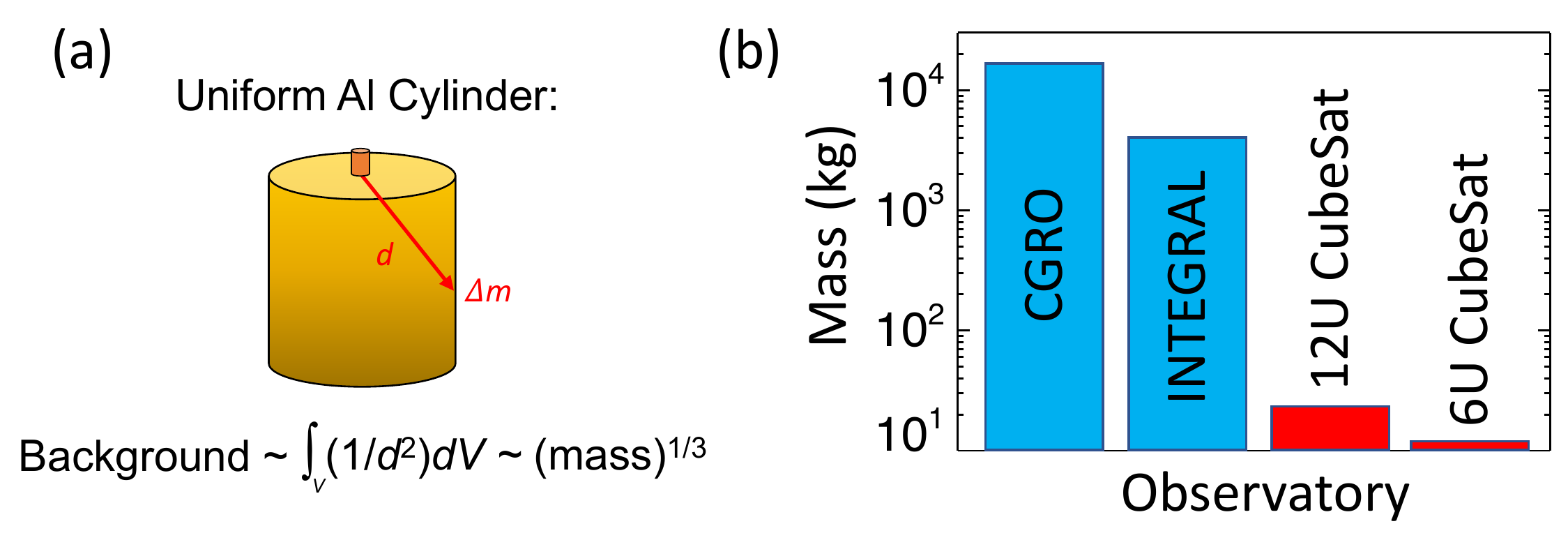}
    \caption{(a) Instrumental background scales approximately as (mass)$^{1/3}$.  (b) CubeSat missions have 2-3 orders of magnitude lower mass than previous missions designed for MeV gamma--ray astronomy.}
    \label{fig_mass}
\end{figure}

A low-mass spacecraft produces less instrumental background than a high-mass one.  The exact scaling is dependent in a complex way on the mass composition and distribution, but a simple estimate can be made by considering uniform cylinders of aluminum, representing the spacecraft bus, with the detector located at the center of one end (Fig.~\ref{fig_mass}a).  Assuming a uniform emissivity due to prompt interactions and activation, the background is then proportional to the integral over the volume of $1/d^{2}$, where $d$ is the distance to the detector.  Performing this calculation for aluminum cylinders, the integrated “background” scales as (mass)$^{1/3}$. 
Comparatively tiny CubeSat platforms therefore offer a uniquely “quiet” environment compared to traditional, large-scale gamma--ray science missions.  As illustrated in Fig.~\ref{fig_mass}b, standard 6U (12\,kg) and 12U (24\,kg) CubeSats have 2-3 orders of magnitude lower mass than previous MeV gamma--ray astronomy missions such as CGRO (17,000\,kg) and INTEGRAL (4,000\,kg), implying roughly an order of magnitude less instrumental background.  Since the gamma-ray detection efficiency for many proposed CubeSat missions is not reduced by nearly this large a factor, a carefully designed small mission can actually provide better observations of relatively bright sources than a large mission for energies around $\sim 1$ MeV.

\subsubsection{Polarimetry}

The polarimetry of high-energy astrophysical sources is a new field that has only begun to be explored.  The polarization of gamma--ray photons provides new insights into physical processes beyond those gleaned from spectroscopy, imaging, or timing.  Much attention has been paid to the potential polarization of the prompt emission from GRBs, which could reveal the structure of the jet magnetic field and the emission mechanism (e.g., synchrotron vs. inverse Compton) \cite{Toma2009, Gill2020}.  Persistent sources such as X-ray binaries \cite{Laurent2011} and gamma--ray pulsars \cite{Harding2019} are also of great interest, as the details of their magnetic field structures and emission mechanisms are unknown and impossible to disentangle through timing and spectroscopy alone.

Since the distribution of the azimuthal scatter angle in Compton scattering is dependent on the polarization of the incident radiation, instruments operating in the soft-to-medium gamma-ray band ($\sim 0.1-10$\,MeV) are inherently sensitive to polarization if they can measure this distribution \cite{Lei1997}.  Since the signal is necessarily divided into multiple angular bins, high signal-to-noise is necessary.  Geometric asymmetries in the instrument can introduce spurious polarization signals, and to mitigate this it is helpful to rotate a Compton polarimeter about its viewing axis; this can be a complex operation for a large spacecraft, but is more feasible for small CubeSats.  Additionally, multiple such polarimeters in a swarm of CubeSats would likely observe a given source with variety of orientations relative to the polarization angle, offering another method of controlling these systematic errors.  

\subsubsection{Nuclear Lines}

Gamma--rays from nuclear processes, such as radioactive decay and de-excitation from nuclear collisions, as well as from electron-positron annihilation, 
often take the form of narrow spectral lines and can provide unique information on the formation and evolution of heavy elements in the Universe \cite{Diehl2022}.  Detection of MeV gamma--ray lines from transient, multi-messenger events such as nearby supernovae or kilonovae will doubtless require large-scale instruments on major missions; we note that nuclear astrophysics is a major motivator of the COSI mission \cite{Beechert2022}.  However, modern, low-resource gamma--ray detectors with very good energy resolution, placed in space on low-background platforms such as CubeSats, may have an important role to play in the study of brighter persistent sources such as the 511 keV electron-positron annihilation emission emanating from the Galaxy \cite{Siegert2016}.

\subsubsection{Cosmic Diffuse Background}

Diffuse cosmic background radiation provides a  constraint on the evolution of stars and galaxies over the history of the Universe.  Although this radiation is well understood across much of the electromagnetic spectrum, the origin of the cosmic diffuse gamma--ray (CDG) background in the MeV energy band is a mystery that has persisted for over 40 years.  It was first measured in the 1970s by simple spectrometers on the Apollo 15 \& 16 Lunar missions \cite{apollo_cdg}.  Massive accreting black holes known as `blazars' are known to emit MeV gamma--rays via particle acceleration and must play a role \cite{Ajello2009}. MeV gamma rays are also produced by nuclear processes in Type Ia supernovae (SNe Ia), and surely also contribute \cite{RuizL2016}.  Existing data, however, are of insufficient quality to precisely discriminate between these (or other) contributions.

The best measurements to date were made by the Compton Telescope (COMPTEL) on NASA’s Compton gamma--ray Observatory (CGRO)\cite{comptel_cdg} and the Solar Maximum Mission (SMM) gamma--ray Spectrometer\cite{smm_cdg}.  Both analyses relied on complex procedures for estimating and subtracting time-variable instrumental backgrounds, both prompt and activation.  Nearly order-of-magnitude systematic error bars are evident in the COMPTEL data, especially below 4\,MeV where activation is most prominent.  In contrast, a CubeSat-based instrument would experience far less instrumental background, and would take advantage of the fact that, since the MeV CDG background is relatively bright, a large detector is not required to make high-quality measurements.  

\section{Currently Operating Gamma--Ray CubeSats}

Several CubeSats aimed at gamma--ray astronomy and related fields have been launched within the past few years, with some still operating.  We review the status of these pioneering missions in this Section.

\subsection{GRBAlpha/VZLUSAT-2}

CAMELOT: Cubesats Applied for MEasuring and LOcalising Transients, is a constellation of 3U CubeSats equipped with large and thin (150 × 75 × 5 mm) CsI(Tl) scintillators read out by SiPM detectors, called multi-pixel photon counters (MPPCs), by Hamamatsu. The detectors are placed on two perpendicular walls of the satellites to maximize the effective photon collecting area on a CubeSat of this size. The detector concept developed for the CAMELOT mission was first demonstrated on a 1U CubeSat, named GRBAlpha. It carries a smaller, 75 × 75 × 5 mm CsI (Tl) scintillator, which provides one-eighth of the expected effective area of the 3U CubeSats envisioned for the CAMELOT mission. 

\begin{figure}
\centering
    \includegraphics[width=1.0\textwidth]{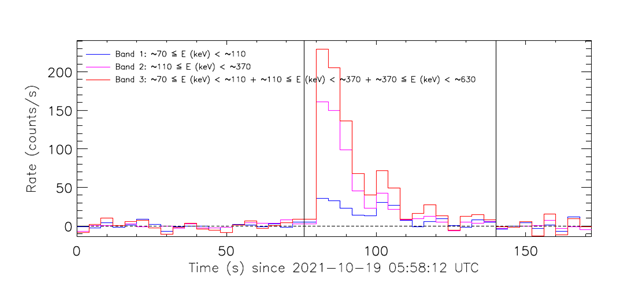}
    \caption{GRB211019A lightcurve obtained by GRBAlpha \cite{GCN30946}}
    \label{GRBAlpha_LC}
\end{figure}

GRBAlpha was launched on 22 March 2021 to a sun-synchronous polar orbit with a Soyuz 2.1 rocket from Baikonur. GRBAlpha is the first 1U satellite to detect multiple confirmed GRBs (Figure~\ref{GRBAlpha_LC}). 

The GRB detectors developed for the CAMELOT mission have been installed on VZLUSAT-2, a technological 3U CubeSat built by the Czech Aerospace Research Centre. The satellite  carries two perpendicular detectors the same size as GRBAlpha. VZLUSAT-2 was released from the D-Orbit ION space-tug on January 26th 2022, following a SpaceX Falcon 9 launch on January 13th. 

GRBAlpha and VZLUSAT-2 are successfully operating in LEO and collecting scientific data \cite{ripa2022}. GRBAlpha has so far detected five GRBs, VZLUSAT-2 has so far detected three GRBs and two solar flares demonstrating that nanosatellites can host payloads sensitive enough to routinely detect these events.
Both satellites are monitoring background in LEO and the maps constructed from these measurements will be useful for the preparation of future GRB missions. The degradation of MPPC sensors on GRBAlpha is being monitored, providing valuable information for future missions which plan to use detectors with MPPC readout.
One year after its launch, the GRBAlpha detector performance is good and the degradation of the MPPC's remains at an acceptable level.

The GRB payload on VZLUSAT-2 has also detected two solar flares to date which coincide with peaks in the X--ray flux detected by the Geostationary Operational Environmental Satellites \cite{ripa2022}. 

\subsection{GRID}

The gamma--ray Integrated Detectors (GRID) mission \cite{grid_wen, grid_zheng} is a concept for continuous, full-sky observations of GRBs using a constellation of 6U CubeSats in low earth orbit (LEO).  Led by Tsinghua University in China, GRID was proposed and developed by students and continues as a student-led project with more than 150 students across 20 Chinese institutions contributing.

Each CubeSat in the GRID constellation carries an identical gamma--ray detector based on Gd$_{3}$(Al,Ga)$_{5}$O$_{12}$ (GAGG) scintillator with SiPM readout, sensitive over an energy range of $\sim 10$ keV -- 2 MeV.  The detector contains a 2 $\times$ 2 array of GAGG crystals, each 3.8 $\times$ 3.8 $\times$ 1 cm$^3$, wrapped in 3M enhanced specular reflector (ESR).  Each of these four scintillators is optically coupled to a 4 $\times$ 4 array of MicroFJ-60035-TSV SiPMs manufacured by SensL/On Semiconductor. The 16 SiPMs for each GAGG crystal are summed into a single readout channel and fed into one of the four readout electronics chains.  An ARM Cortex M0+ micro-controller controls the readout and communicates with the CubeSat bus.  No on-board temperature or gain control is used, but the temperature of each SiPM array is monitored with a digital temperature sensor for use in data analysis.  

The first GRID detector was launched on 29 October 2018 on a 6U CubeSat developed by Spacety Co. Ltd. into a Sun-synchronous orbit with an altitude of 500 km and inclination of 97.5$^{\circ}$.  It accumulated data for approximately one month after its on-orbit functional and performance tests were concluded in 2019.  This first prototype was followed into space by GRID-02 (6 November 2020) and GRID-03 \& GRID-04 (27 February 2022), all launched into polar orbits with inclinations of $\sim 97^{\circ}$ and altitudes between 450 and 550 km.  To date numerous GRBs have been detected (e.g., \cite{grid_grb}).  In addition, in GRID-02 a gradual increase in the SiPM dark current of $\sim$ 100 $\mu$A per year per chip has been reported \cite{grid_zheng}, providing valuable  data on long-term SiPM behavior in the orbital radiation environment.
 
\subsection{LIGHT-1}
\begin{figure}
\centering
    \includegraphics[width=0.9\textwidth]{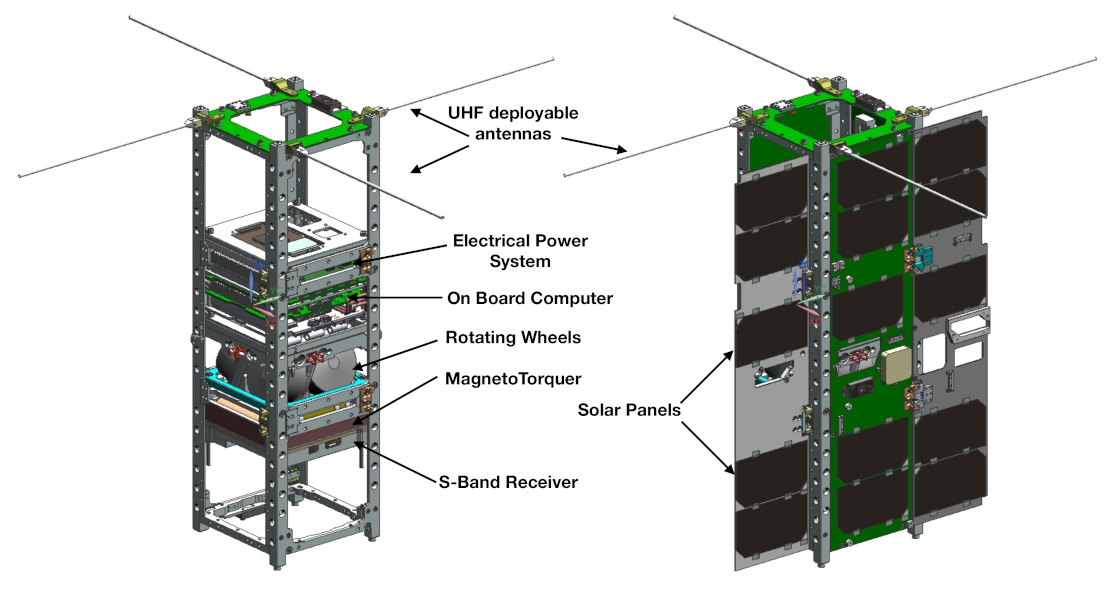}
    \caption{CAD view of the 3U LIGHT-1 CubeSat mission showing (left) the main subsystems. These fit within 1.3\,U,
leaving 1.7\,U for the 2 gamma--ray detector payloads at the top and bottom of the spacecraft. Deployable and embedded
solar panels used for power generation are shown on the right \cite{Arneodo2021}}
    \label{Light1}
\end{figure}

Light-1 (Figure~\ref{Light1}) is a 3U CubeSat that was deployed from the ISS in 2022 and whose primary mission is the study of TGFs. The mission is based on a collaboration between Khalifa University of Science and Technology and New York University Abu Dhabi and has been implemented mainly by students \cite{Almazrouei2021}. There are 2 gamma--ray detectors, each occupying $\sim$\,1U volume. Given the fast temporal variability of TGFs and the high count rates they generate, the fast decay scintillating material,   cerium bromide (CeBr3(LB)) is used \cite{Iyudin2022}.  One scintillator is coupled to a Hamamatsu R11265U-200 photomultiplier and the other to an SiPM array \cite{DiGiovanni2019}. 
\subsection{MinXSS}

In soft X--rays, the goal of the Miniature X--ray Solar Spectrometer' (MinXSS) suite of 3U CubeSats is the study of processes in the dynamic Sun, from quiet-Sun to solar flares \cite{Mason2016,Mason2020}. The enabling technology is the Amptek X123, a commercial-off-the-shelf (COTS) silicon drift detector (SDD), with low mass, modest power consumption, and small volume, making it suitable for use on a CubeSat. 

MIN-XSS1 operated successfully for 1 year. Although MIN-XSS2 was launched in December 2018, contact with the spacecraft was lost in January 2019, as a result of a communications anomaly that occurred in response to a single-event latchup in the command and data handling unit   
\footnote{\url{https://lasp.colorado.edu/home/minxss/2019/02/14/minxss-2-communications-anomaly/}}.

\section{Gamma--Ray CubeSat Missions Under Development}

In this Section we review gamma--ray CubeSat missions that are funded and in an advanced state of development, with planned launches within the next several years.

\subsection{BurstCube}

BurstCube is a 6U CubeSat designed to detect and localize GRBs in the 50\,keV to 1\,MeV energy range and then quickly downlink the information via the NASA Space Network (SN) \cite{perkins2020}. BurstCube is funded via NASA's APRA program.  Rapidly communicating information about GRBs to the astrophysical community is one of the cornerstones of GRB science, as demonstrated by missions like Swift and Fermi, and it is a clear challenge for smaller missions due to power and cost constraints.  BurstCube is a pathfinder in using the SN on a small platform.  The communications are provided by a commercial software defined Vulcan NSR-SDR-S/S S-band radio that connects to the Near Earth Network and the SN.  The radio will transmit at high power (needed for the SN link) and low power to the ground.  

\begin{figure}
\centering
    \includegraphics[width=0.5\textwidth]{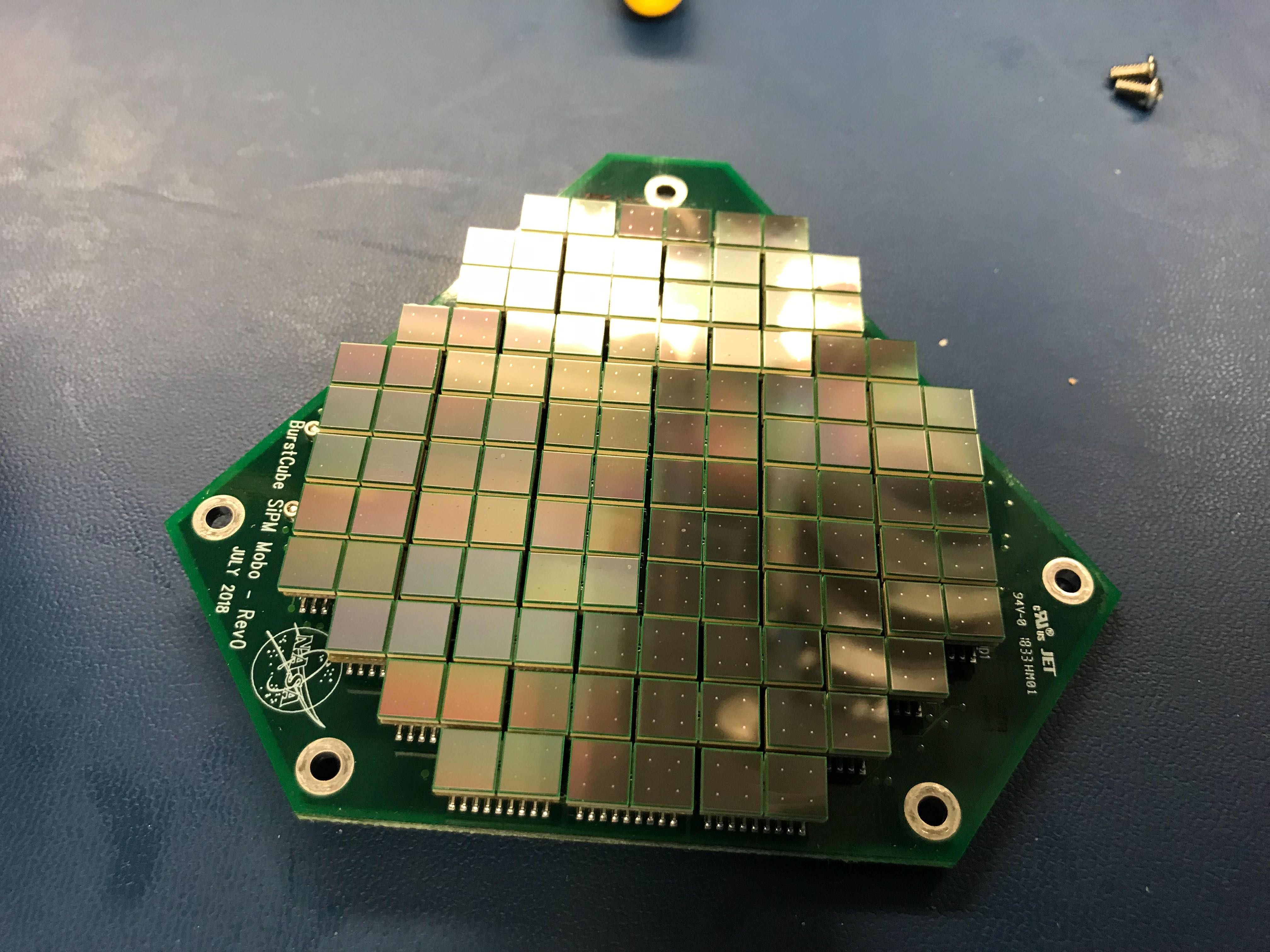}
    \caption{The BurstCube SiPM arrays are each composed of 116 SiPMs (Hamamatsu 13360-6050 multi-pixel photon counter) whose signals are summed into a single analog output per CsI crystal.}
    \label{BurstCube_SiPMs}
\end{figure}

The BurstCube instrument is composed of four 90 mm diameter cylindrical CsI crystals which are 19 mm thick.  These are viewed by arrays of SiPMs (Figure \ref{BurstCube_SiPMs}).  
Each detector is oriented approximately 45 degrees from zenith giving a good field of view and reasonable localizations.  The signals from the SiPMs are summed in banks (to reduce the overall capacitance load on the front end) and digitized continuously in an FPGA which also does peak finding.  These peaks are passed to the spacecraft FPGA where binning, burst detection and data handling occur.  

\begin{figure}
\centering
    \includegraphics[width=0.5\textwidth]{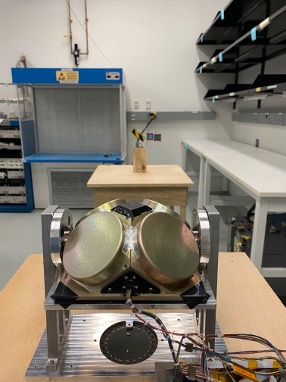}
    \caption{The BurstCube instrument is fully assembled and tested. The assembled instrument is shown here in a test stand during instrument calibrations.}
    \label{BurstCube}
\end{figure}

\begin{figure}
\centering
    \includegraphics[width=0.7\textwidth]{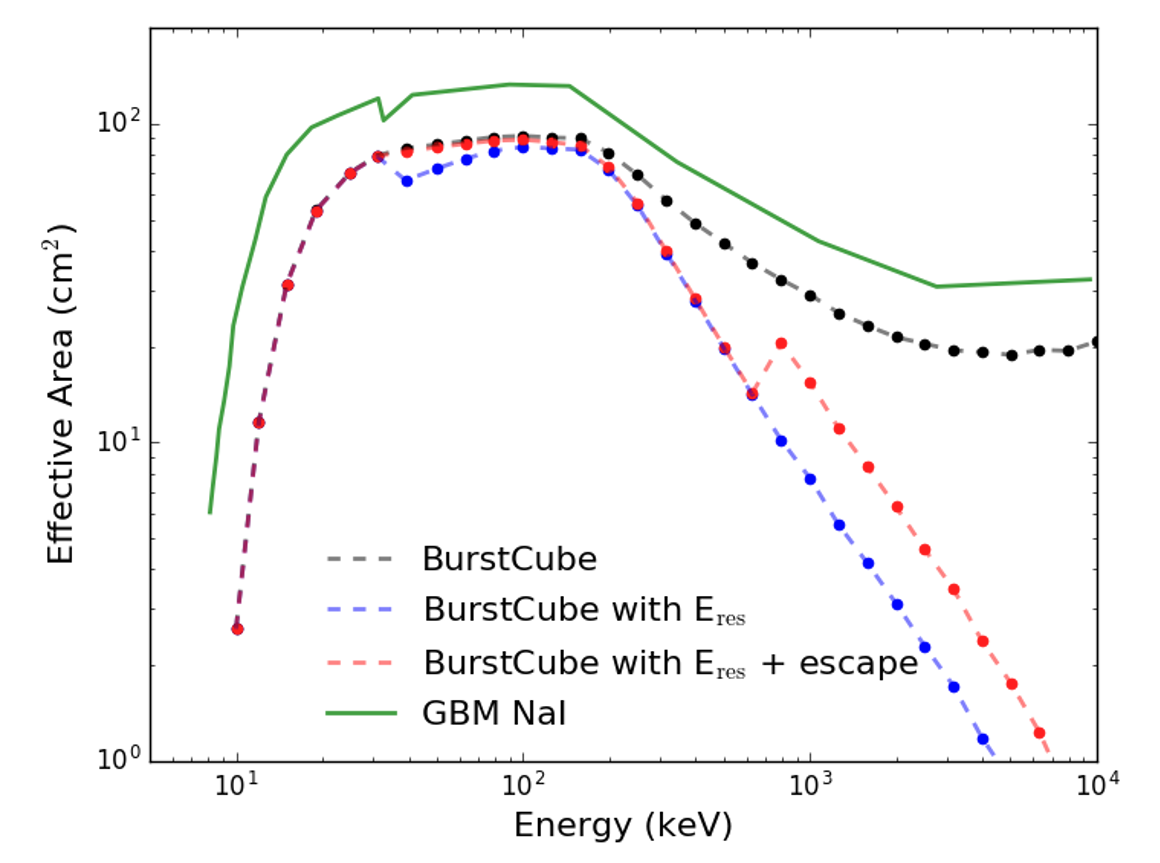}
    \caption{The BurstCube instrument achieves an effective area within a factor of $\sim 2$ of that of the Fermi/GBM NaI detector, while fitting within a 6U CubeSat.}
    \label{BurstCube_EA}
\end{figure}

The BurstCube instrument has undergone thorough testing and calibration (Figure \ref{BurstCube}).  The effective area of the instrument is shown in Figure \ref{BurstCube_EA} compared to that of the Fermi Gamma-ray Burst Monitor (GBM) \cite{Meegan2009} sodium iodide (NaI) detectors; a comparable efficiency is achieved despite the much smaller payload and mission parameters. The delivery of BurstCube to NanoRacks is scheduled for late 2022 with deployment from the ISS in late 2022 or early 2023.  


The BurstCube team has made significant effort in releasing software (simulations, science analysis, and data pipeline) as open source and freely available.  All of the tools have been developed with multiple mission analysis in mind and are usually for general use.  Budgets for small missions are usually constrained and the reuse of software is one way missions can collaborate and develop the field together. The software has a friendly API and can be used by other CubeSats or SmallSat missions with similar instruments and data analysis chains. The software (bc-tools) relies on core functionality delivered by the gbm-data-tools which were developed by the GBM team showing that it's also important to rely on the support of larger missions.  As bc-tools have been developed, functionalities and algorithms that are of more general use are incorporated into gbm-data-tools by the bc-tools developers.  

In addition to software, the BurstCube team has a commitment to release data products to the community as rapidly as possible.  There will be several products available.  The base data set on board are time tagged events (TTE, pulse heights and times).  The on board software searches this data set for periods of significantly high rates and telemeters a portion of the TTE to the ground around triggers and sends out notifications via the SN.  The TTE is also stored on board for a period of time.  Users on the ground can ask for portions of the TTE data (Requested TTE or RTTE) to be sent to the ground at the next data pass.  Finally, the onboard software bins the TTE in energy and time to send to the ground and continuously binned data (CBD) for ground searches and analysis.      

\subsection{EIRSAT-1}
\begin{figure}
\centering
\includegraphics[width=0.8\textwidth]{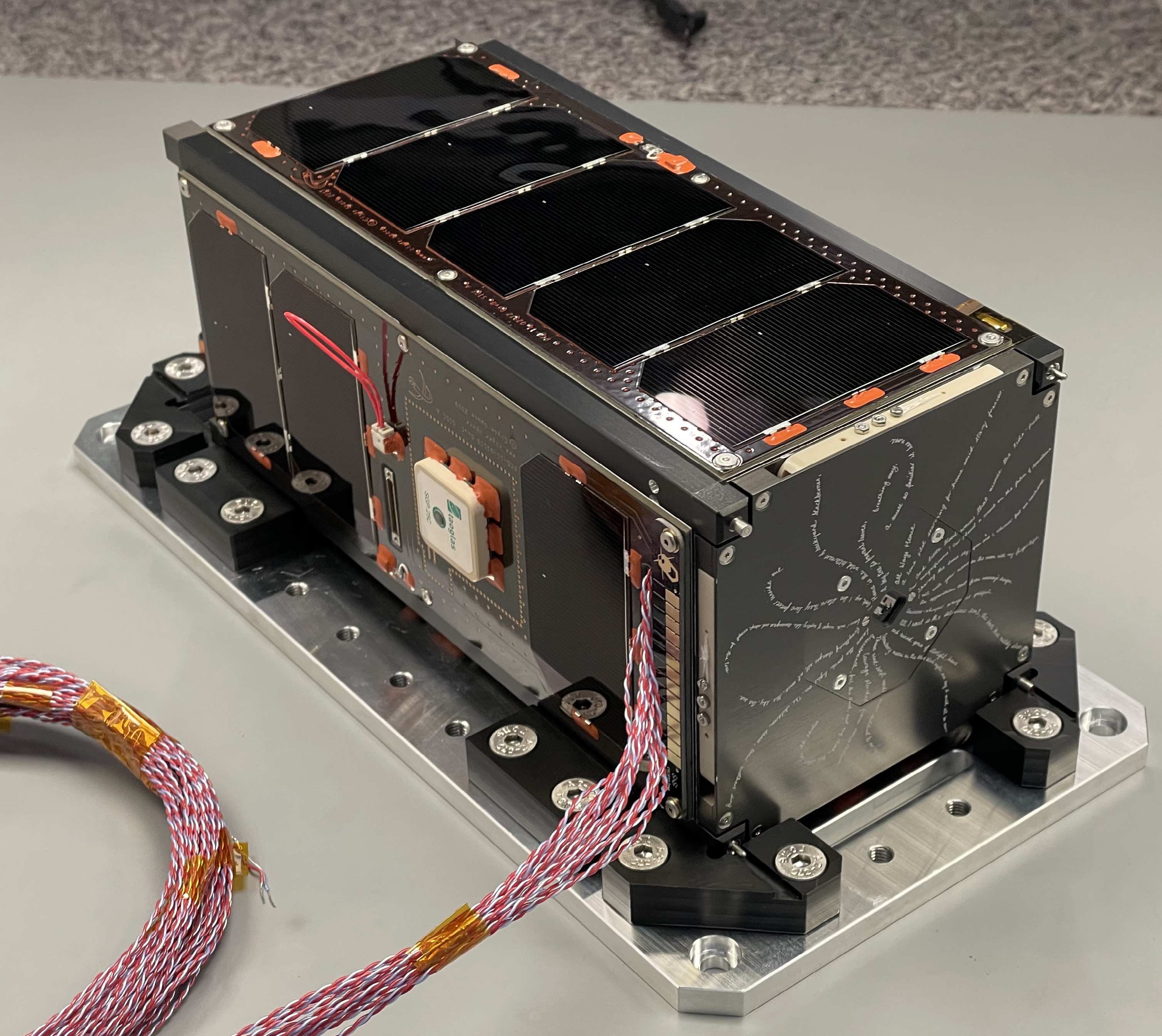}
\caption{EIRSAT-1 Flight Model. Credit: J. Thompson}
\label{fig:eirsat1}
\end{figure}

The Educational Irish Research Satellite 1 (EIRSAT-1, Figure~\ref{fig:eirsat1}) is a 2U CubeSat which is Ireland's first satellite~\cite{Doyle2022a}. EIRSAT-1 was proposed in response to an ESA announcement of opportunity for their educational Fly Your Satellite! programme and was accepted into that programme in 2017. The main objective of the mission is to build capacity in the Irish higher education sector in space science and engineering with three novel payloads developed in-house. The ENBIO Module (EMOD) is a thermal materials experiment; Wave-Based Control (WBC) is a software-based attitude control test-bed \cite{Sherwin2018}, and the Gamma--ray Module (GMOD), discussed below, is a compact $\gamma$--ray instrument for GRB and other high-energy transient source detection. An antenna deployment module, responsible for deploying the mission's UHF/VHF antenna elements on-orbit, was also developed in-house \cite{Thompson2022}. The remainder of the spacecraft consists of COTS components supplied by AAC Clyde Space\footnote{\url{https://www.aac-clyde.space/}}, including the battery, electrical power system (EPS), solar cells, radio transceiver, attitude determine and control system (ADCS), and on-board computer (OBC). The electronic and mechanical interfaces are compatible with many off-the-shelf CubeSat systems and structures. 
In view of its educational goals, many aspects of the mission's development have been disseminated, including software \cite{Doyle2022b}, AIV \cite{Walsh2021}, firmware \cite{Mangan2022}, testing \cite{Dunwoody2022a,Doyle2022c,Reilly2022} and operations \cite{Dunwoody2022b}.

\subsubsection{Gamma--Ray Module (GMOD)}
GMOD evolved from a prototype building block module for the pixellated calorimeter of a full Compton-pair telescope that used a monolithic LaBr$_3$(Ce) crystal scintillator with an SiPM array, read out by a low--power ASIC called `SIPHRA' \cite{Meier2016}. Designed using this prototype SiPM array as a reference, SIPHRA was developed by Integrated Detector Electronics AS (IDEAS) as a general purpose readout IC for photon detectors. The LaBr$_3$(Ce) scintillator was replaced by CeBr$_3$, mainly due to its lower intrinsic background. The module was flown on a high-altitude balloon \cite{murphy2021b} and tested under proton irradiation on the ground  \cite{ulyanov2020}. 

\begin{figure}
\centering
\includegraphics[width=0.9\textwidth]{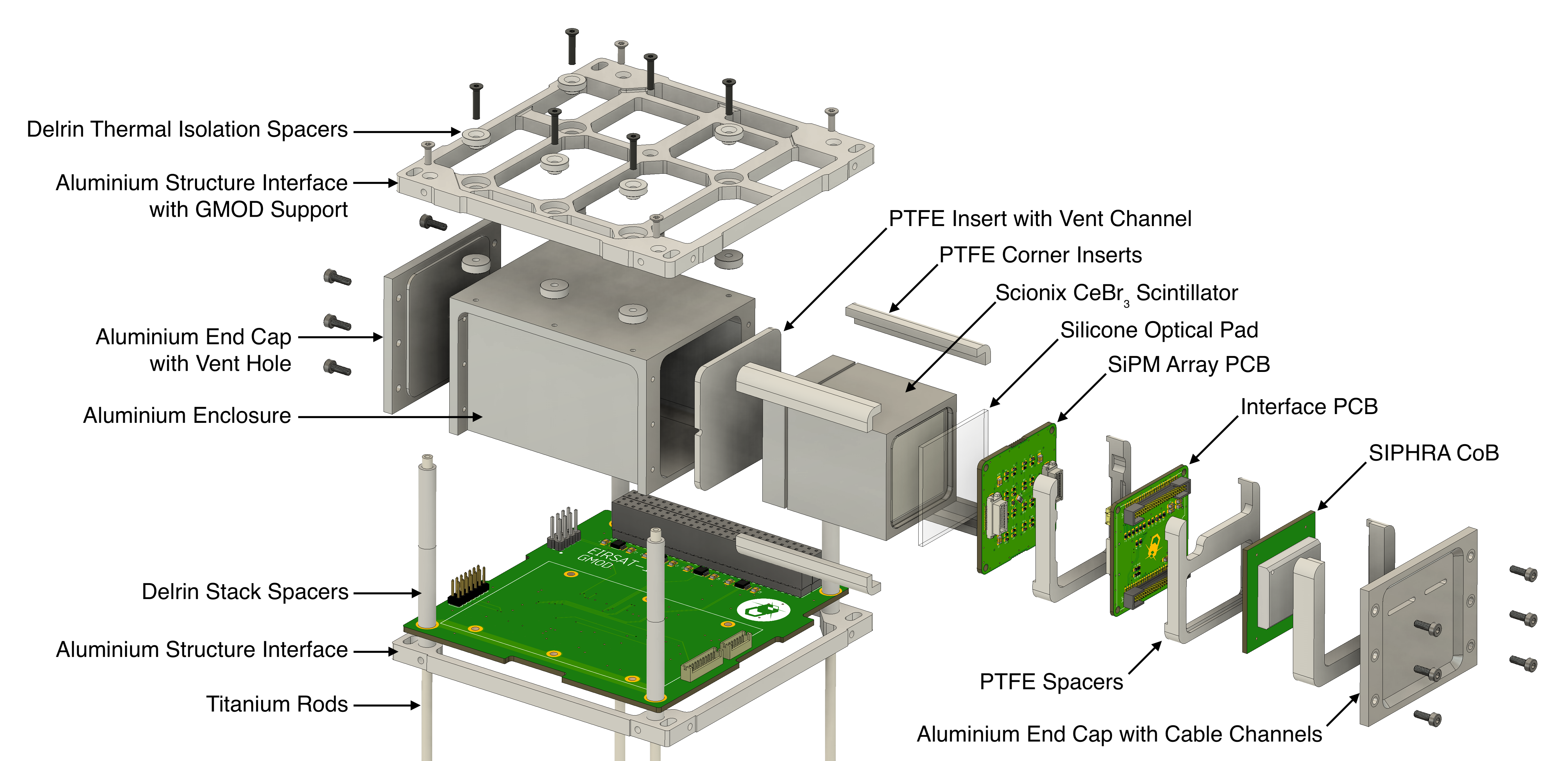}
\caption{Exploded view of GMOD on EIRSAT-1, showing the mounting bracket, detector assembly and CubeSat-compatible motherboard}
\label{fig:explodedview}
\end{figure}

The detector assembly and the motherboard are the two main elements of GMOD (Figure~\ref{fig:explodedview}). GMOD uses a 25\,$\times$\,25\,$\times$\,40\,mm\,$^{\rm 3}$ cerium bromide (CeBr$_3$) crystal scintillator supplied by Scionix enclosed within a hermetically sealed unit. To measure the scintillation light produced by the crystal, GMOD uses sixteen J-series 60035 SiPMs produced by SensL (now ON Semiconductor). The SiPM breakdown voltage varies with temperature \footnote{\url{https://www.onsemi.com/pdf/datasheet/microj-series-d.pdf}}. As SiPM gain is a function of the over-voltage above breakdown at which the SiPMs are biased, this temperature dependence of the breakdown voltage leads to an overall temperature dependence for the instrument calibration. The reverse side of the SiPM array therefore contains a PT100 temperature sensor placed at the centre of the board to measure the SiPM temperature. This sensor is digitised by the SIPHRA ASIC each time the SiPM outputs are sampled, allowing for periodic adjustments to be made to the bias voltage or for appropriate temperature-dependent calibration to be applied on the ground \cite{Ulyanov2017}. 

The GMOD Motherboard is a CubeSat PC-104 compatible PCB which hosts the support electronics necessary to operate the detector assembly. It includes a micro-controller, responsible for data collection, and a CPLD for interfacing to SIPHRA's serial data output. An SPI flash memory allows the data to be cached in GMOD before being sent to the spacecraft on-board computer. The motherboard also hosts the power conditioning for the detector assembly including the adjustable bias PSU which generates the negative bias voltage required by the SIPM array. The motherboard conforms to the de facto standard for CubeSat components popularised by the Pumpkin Inc. CubeSat Kit, measuring 90.17\,mm\,$\times$\,95.89\,mm and interfacing to EIRSAT-1's main system bus through a pair of Samtec ESQ-126-39-G-D stack-through headers.

The GMOD component of the on-board software processes summed TTE data to generate light-curves, spectra, and trigger data-types. The trigger data is incorporated into EIRSAT-1's beacon for immediate broadcast.
The detector's performance characteristics are shown in Table~\ref{gmod:table}. In addition to the educational and technology demonstration goals of the EIRSAT-1 mission, GMOD is expected to detect between 11 and 14 GRBs per year with significance greater than 10$\sigma$ \cite{Murphy2021}. 

\begin{table}
\caption{GMOD detector key parameters}
\label{gmod:table} 
\begin{tabular}{l l}
\hline\noalign{\smallskip}
Energy Resolution & 5.4\% ($@$662\,keV) \\
Effective Area            & 10\,cm$^2$ (sky-averaged) \\
Energy Range              & 30\,keV--2\,MeV  \\
Detector Assembly Mass    & 336\,g  \\
Detector Assembly Volume  & 160\,cm$^3$ \\
Motherboard Mass          & 54\,g  \\
Motherboard Size          & 90.2\,mm\,$\times$\,95.9\,mm \\
Power Consumption         & $\approx$400\,mW \\
\noalign{\smallskip}\hline
\end{tabular}
\end{table}

\subsection{HERMES-Pathfinder}

HERMES-Pathfinder is a constellation of six 3U nano-satellites hosting simple but innovative X-ray detectors for the monitoring of cosmic high-energy transients and for the determination of their position, \cite{fiore, evangelista,sanna}. The  HERMES Technological Pathfinder project is funded by the Italian Space Agency while the HERMES Scientific Pathfinder project is funded by the European Union's Horizon 2020 Research and Innovation Programme under Grant Agreement No. 821896.  HERMES-Pathfinder is an in-orbit demonstration, that should be tested in a low inclination ($i < 20^{\circ}$) LEO orbit starting from 2023.

 In the next few years Advanced LIGO/VIRGO, KAGRA in Japan and LIGO/India will reach their nominal sensitivity. In the electromagnetic domain the Vera C. Rubin Observatory will come on line in 2023, revolutionizing the investigation of transient and variable cosmic sources in the optical band. The operation of an efficient X-ray all-sky monitor with good localisation capabilities will have a pivotal role in providing the high-energy counterparts of gravitational wave interferometers and the Vera Rubin Observatory, bringing multi-messenger astrophysics to maturity. To gain the required precision in localisation and timeliness for an unpredictable physical event in time and space requires a sensor distribution covering the full sky. The main objective of HERMES-Pathfinder is to demonstrate that accurate localisations of high-energy cosmic transients can be obtained using miniaturized hardware. The transient position is obtained by studying the difference in the arrival time of the signal to different detectors hosted by nano-satellites in low Earth orbits \cite{sanna}. Particular attention is paid to reaching the best time resolution and time accuracy, with the goal of reaching an overall accuracy of a fraction of a micro-second. The HERMES pathfinder detector system includes 60 GAGG scintillator crystals and 12 10×10 silicon drift detector mosaics used to read out the crystals \cite{evangelista}. The consortium started the integration and testing of the first flight unit during the summer of 2021; the proto-flight model and its qualification review is foreseen for Q2 2022. The other five units will be integrated and tested during 2022 and the constellation is set to be launched to a nearly equatorial LEO in 2023. SpIRIT (Space Industry - Responsive - Intelligent – Thermal) is a 6U CubeSat funded by the Australian Space Agency and managed by the University of Melbourne. SpIRIT will host one HERMES pathfinder payload, and will fly on an SSO at the same time as the HERMES pathfinder, forming a constellation of seven satellites in two different orbits, thus greatly increasing the localization capabilities for cosmic transients.

\subsection{MAMBO}
The Mini Astrophysical MeV Background Observatory (MAMBO) is a CubeSat mission under development at Los Alamos National Laboratory \cite{mambo_ves, mambo_blos1, mambo_blos2}.  Launch is currently planned for the fall of 2023.  Unlike the missions described previously, MAMBO is not dedicated to the study of bright transient objects, and is designed to achieve good sensitivity above energies of 1 MeV.  Taking advantage of the low instrumental background afforded by a 12U CubeSat bus, MAMBO will record the best measurement ever made of the CDG background in the 0.3 – 10 MeV energy range.  The project thus leverages the unique characteristics and low costs of small missions for a specific scientific goal.  

\begin{figure}
    \centering
    \includegraphics[width=0.8\textwidth]{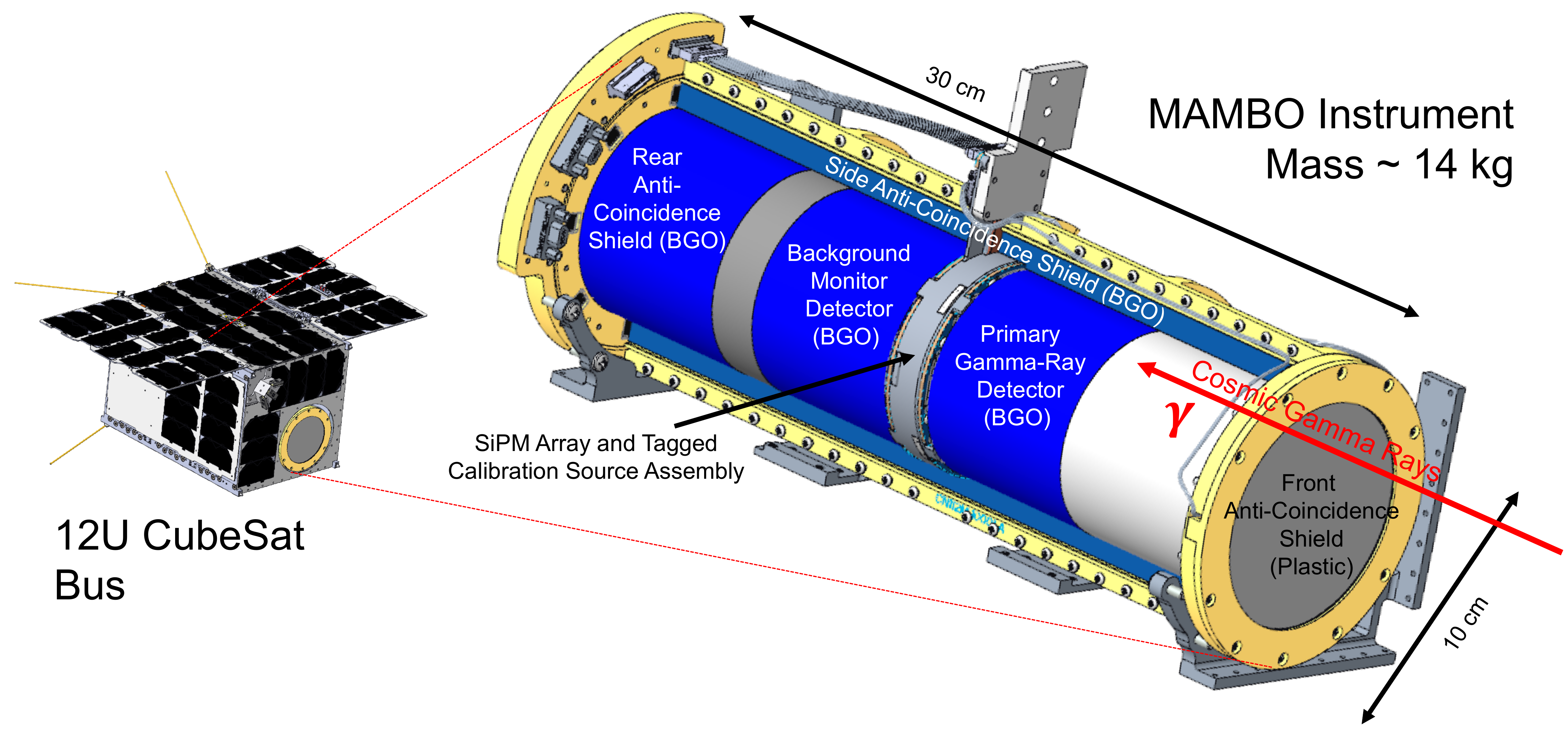}
    \caption{The MAMBO instrument achieves high efficiency in a compact configuration suited for deployment on a 12U CubeSat bus.}
    \label{fig_mambo_overview}
\end{figure}

\begin{figure}
    \centering
    \includegraphics[width=0.8\textwidth]{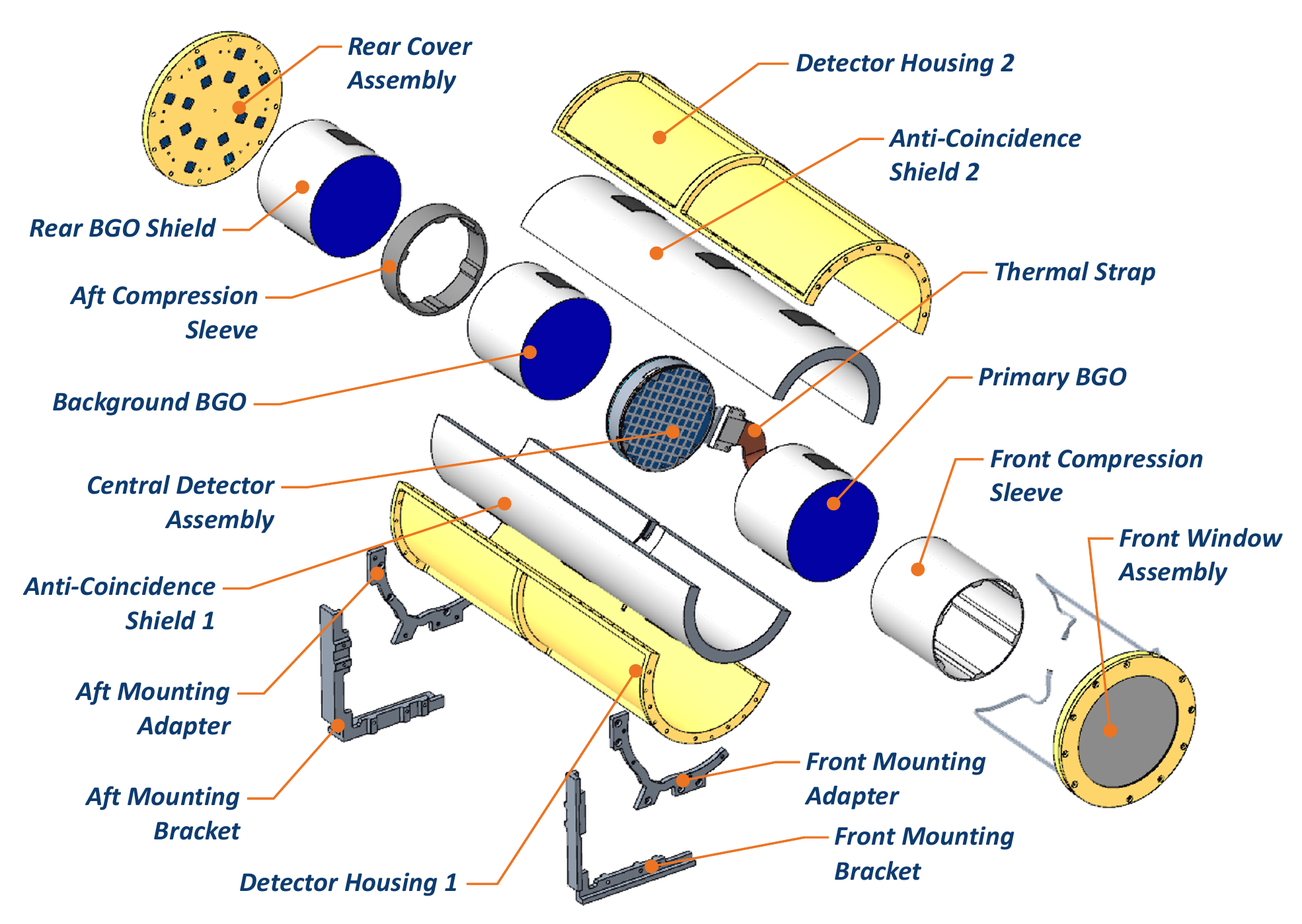}
    \caption{Exploded view of the MAMBO instrument components.}
    \label{fig_mambo_exp}
\end{figure}

The MAMBO instrument is illustrated in Figs.~\ref{fig_mambo_overview} and \ref{fig_mambo_exp}.  MAMBO achieves high efficiency and exceptional background rejection using an innovative shielding configuration that simultaneously measures signal and background.  
The MAMBO spectrometer employs two identical, closely-spaced Bismuth Germanate (BGO) scintillator detectors.
The front scintillator, denoted the Primary Detector, is a 3$''$ (diameter) $\times$ 2.5$''$ (length) cylinder of BGO and is exposed to the CDG background through the entrance aperture (Fig.~\ref{fig_mambo_overview}).  The rear, identical, BGO is designated the Background Monitor and is shielded from the CDG background by the Primary Detector.  Both detectors are exposed to the same instrumental background from the sides.  Subtraction of “Monitor” spectra from “Primary” spectra thus removes the instrumental background even during fast variations, a fundamentally new approach to the traditional practice of calculating or simulating the background.  Cosmic radiation incident from the sides is subtracted as well, suppressing the response to off-axis sources.  Sources of radiation outside the field of view are also suppressed by the surrounding Side and Rear BGO anti-coincidence shields. An anti-coincidence plastic scintillator shield over the front aperture rejects charged particles while allowing gamma rays to enter.

\begin{figure}
    \centering
    \includegraphics[width=0.3\textwidth]{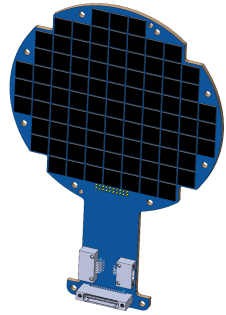}
    \caption{The MAMBO SiPM array containing 86 J-series SiPMs from SensL/On Semiconductor.}
    \label{fig_mambo_sipms}
\end{figure}

To enable CubeSat deployment, the scintillator light output is measured by SiPMs (Figs.~\ref{fig_mambo_exp} and \ref{fig_mambo_sipms}).  
The Primary Detector and Background Monitor are read by identical, custom arrays of SiPMs which are designed for maximum compactness and sum the individual array elements into a single output signal for each scintillator.  The MICROFJ-60035-TSV device, a 6 mm $\times$ 6 mm “J-series” SiPM manufactured by SensL/ON Semiconductor, is used to construct these arrays. The anti-coincidence shields are read out by individual SiPMs mounted on small printed circuit boards (PCBs) in the case of the plastic shield, or by summed SiPMs on a “Rear PCB” for the Side and Rear BGO Shields; these signals “veto” the processing of coincident Primary Detector or Background Monitor events.  As SiPM gain is dependent on temperature, the readout system will monitor thermistors and coarsely adjust the bias voltage of the shield SiPMs in order to maintain an approximately constant gain.
It is critical that the gains of the Primary Detector and Background Monitor be kept precisely matched in order for the MAMBO background subtraction technique to work.  This is accomplished using a “tagged” calibration source: a small button of plastic scintillator infused with $^{60}$Co, with its own dedicated SiPM readout, produces a flash of light due to the $\beta$-particle emitted in coincidence with 1.17 MeV and 1.33 MeV gamma rays.  This identifies calibration events in each of the two main detectors; these events are analyzed by an algorithm running on a Field Programmable Gate Array (FPGA) to determine necessary bias voltage adjustments to maintain matched and constant gains. The SiPM Bias Voltage Board generates an adjustable (25\,V--31\,V) bias voltage for each of the SiPM readouts in the instrument. The bias levels for the shield and tagged $^{60}$Co source SiPMs will be adjusted according to the temperature via look-up tables. 

The MAMBO payload will be flown on a N12P 12U bus provided by NanoAvionics Corp\footnote{\url{https://nanoavionics.com/}}.  The standard N12P bus can meet all of the MAMBO mission requirements with only minor custom modifications.  The launch, currently planned for the fall of 2023, will be provided by the U.S. Department of Defense STP.  The ideal science orbit would be an equatorial low Earth orbit (LEO), but readily available options through STP are more likely be be at an inclination of $\sim 55^{\circ}$; the MAMBO background subtraction technique will permit sensitive observations of the MeV CDG background in this orbit as well.  MAMBO will map the sky from 0.3 – 10 MeV via a series of pointed observations lasting $\sim 10^6$ seconds each.  Time-variable backgrounds will be extrapolated to zero to reveal the residual constant signal from the CDG background, following and building on the analyses of COMPTEL and SMM.  The instrument field of view is $\sim 1$ steradian.  Preliminary background simulations  predict that the total counting rates in the two main BGO detectors should never exceed 500 cts s$^{-1}$ outside of the radiation belts, yielding a data volume of $< 600$ MB per day.  MAMBO will use a commercial ground station network for telemetry and commanding.

\subsection{IMPRESS}
The IMPRESS CubeSat is being developed for solar flare observations at the University of Minnesota SmallSat Research Laboratory, with Montana State University, the University of California, Santa Cruz and the Southwest Research Institute (SwRI). The goal is to measure HXR spikes on timescales of tens of milliseconds. It features four custom cerium bromide scintillator detectors with SIPMs, sensitive in the hard X-ray/soft gamma--ray band of 10--300\,keV and read out by Bridgeport Instruments electronics, along with an  AmpTek X-123 FastSDD, sensitive in the range 1--12\,keV \cite{Setterberg2022}. The scintillator-based detector will investigate flare nonthermal bremsstrahlung emission, while the X--ray detector will constrain thermal bremsstrahlung emission. Solar flare HXR emission will be recorded at a frame rate/cadence of 32\,Hz to measure fast HXR spikes and thereby constrain acceleration mechanisms, and the SXR emission at a 1\,Hz frame rate. A secondary mission goal is to perform stereoscopic X--ray directivity measurements in conjunction with the STIX instrument onboard Solar Orbiter to establish the nature of beaming in the non-thermal emission of solar flares.

\subsection{LECX}
As an alternative to the more common scintillator-based CubeSat instruments for the study of high-energy transients, the Localizador de Explos\~{o}es C\'{o}smicas de Raios X (LECX – Portuguese for Locator of X-ray Cosmic Explosions) uses CZT detectors to detect and localize GRBs in the $30-200$\,keV energy band \cite{lecx}.  The experiment consists of an array of four planar CZT detectors, each $10 \times 10 \times 2$ mm$^3$, surrounded by a passive shield of graded Pb-Sn-Cu to reduce background and define a $53^{\circ} \times 53^{\circ}$ field of view.  The LECX experiment is scheduled to fly on the nanoMIRAX satellite, a 2U CubeSat developed by the Brazillian private sector, with an expected launch in 2023.  LECX represents the use of advanced compact detector technology to create very low Size, Weight and Power (SWAP)  solutions suitable for flight on large swarms of smaller CubeSats.

\section{Other Proposed Gamma--Ray CubeSat Concepts}

In addition to the operational missions and advanced concepts described above, many other conceptual applications of CubeSats to gamma--ray astronomy have been proposed in the literature.  We describe several that span a wide range of scientific goals and technological approaches.  We note that this field is quite active and dynamic, and so the survey presented here is undoubtedly far from comprehensive.

\subsection{CubeSats for Bright Transients}

The gamma--ray burst Localizing Instrument (GALI) is a concept for a compact GRB detector with enhanced localization capabilities \cite{gali}.  GALI employs a three-dimensional array of small (9-mm cubic) scintillators, currently CsI(Tl), with SiPM readouts. The scintillators are arranged such that they mutually occult each other with different shadowing patterns for different incident angles.  A given distribution of scintillator count rates thus corresponds to a unique GRB location on the sky.  GALI is not currently planned for a specific mission, but the existing prototype instrument is sized to fit approximately within a 1U volume, and the concept is easily scalable to various CubeSat or other small satellite formats.  Such an instrument concept, as is the case with many of the missions described here, is enabled by the combination of advanced SiPM and ASIC technology.

A constellation of 6-8 nanosatellites for TGFs, called `BEES', is currently in Phase A study by CNES as a replacement for the failed TARANIS mission.  Launch is planned for around 2030. BEES will use 16\,GAGG\,+\,SIPM pixels, read out by the IDEAS/APOCAT ASIC. 

\subsection{CubeSats for Gamma--Ray Polarimetry}

Taking advantage of low-SWAP, multi-channel detector and electronics technology, CubeSats are being developed to make pioneering measurements in the relatively new field of gamma--ray polarimetry.  The COMCUBE concept \cite{laviron2021}, funded by the European AHEAD2020 program, seeks to place a constellation of 6U CubeSats into LEO carrying wide-field-of-view Compton polarimeters to measure the polarization of the prompt emission from GRBs.  The proposed instrument consists of two scattering layers, each containing four double-sided silicon strip detectors, and a number of calorimeter modules, each composed of a CeBr$_3$ scintillator crystal with a SiPM array readout, placed below and to the sides.  The instrument is designed to fit within 4U of a 6U CubeSat volume and would provide imaging and spectroscopy as well as polarimetry of GRBs.  Initial simulation studies indicate that a constellation of four COMCUBE CubeSats would measure about five GRBs per year with a minimum detectable polarization (MDP) of $<$ 30\% at the 99\% confidence level.  

Similar technology is used, with a different approach for a different scientific goal, in the COMPOL concept \cite{Yang2020, laurent2022}.  A small Compton polarimeter with a narrow field of view, flown on a 3U CubeSat, is designed to measure the polarization of Cygnus X-1, a bright, persistent source.  The small detection efficiency of the instrument is compensated for by a very long observation time, as the mission is dedicated to observing only this one source for a period of a year or more.  The scattering element will be an array of silicon drift diode "pixels," and the calorimeter a single large CeBr$_3$ crystal read out by an array of SiPMs to derive position information by the distribution of scintillation light (a so-called ``Anger camera'').  Simulations indicate that COMPOL will achieve an MDP of $\sim 18$\% in the 10 -- 300 keV energy band in a six-month observation.  An in-orbit validation model, ComPol-ISS, will be launched to the International Space Station in 2023, and the full 3U CubeSat mission is planned for launch in 2026.

\subsection{CubeSats for Gamma--Ray Line Studies}

The inherently low instrumental background afforded by a low-mass CubeSat platform opens new possibilities for the study of gamma--ray lines, a key technique for nuclear astrophysics.  One such proposal \cite{Hughes2022} focuses on the 511 keV gamma--ray line from electron-positron annihilation.  An array of seven $56 \times 53$ mm cylindrical high-purity Germanium (HPGe) crystals would fit within the volume of a 12U CubeSat, and provide the best energy resolution available at gamma--ray energies ($\sim 0.3$\% FWHM at 511 keV) provided they could be cooled to $< 100$ K.  Off-the-shelf cooling technologies compatible with a 12U CubeSat SWAP are available, such as the Northrup-Grumman Space Micro Pulse Tube Cooler cryocooler.  Such an instrument flown on a 12U bus would achieve an active detector mass fraction of over 30\%, greatly improving sensitivity compared to the SPI instrument on INTEGRAL (mass fraction $\sim 0.6$\%).  Mapping the 511 keV intensity on the sky could be achieved using the Earth occultation technique.

Similar arguments motivate a proposal for a ``Lunar CubeSat'' mission concept \cite{Pinilla2020} for high-sensitivity nuclear astrophysics.  To realize very good energy resolution ($< 1$\% FWHM at 662\,keV) without the need for cryogenic cooling, this concept employs thick, pixellated CZT detectors flown on a 12U CubeSat.  The spacecraft would be placed in orbit around the Moon and utilize the lunar occultation technique \cite{Miller2016} to achieve high sensitivity and angular resolution by analyzing the changes in flux as gamma--ray sources rise and set behind the sharp lunar limb.  Preliminary sensitivity studies for a mission carrying 32 CZT detectors, each $4 \times 4 \times 1.5$ cm$^3$, indicate narrow-line sensitivities an order of magnitude better than previous missions, especially at 511\,keV due to the lack of the strong atmospheric line encountered in LEO.

\subsection{CubeSats for General MeV Astrophysics}

Gamma--ray astronomy in the MeV energy band has long struggled to achieve sensitivity comparable to that at other wavelengths.  The MeVCube concept \cite{Lucchetta2022} seeks to address this need, utilizing advanced detector technology and a low-mass CubeSat platform to realize good energy and angular resolution for a wide range of science goals in the 0.2--4\,MeV energy band, beyond bright transients or nuclear lines.  The instrument concept consists a Compton telescope comprising two layers of 64 CZT detectors, each $2 \times 2 \times 1.5$ cm$^3$, with a separation of 6 cm.  Each CZT detector has an $8 \times 8$ array of pixels on a 2.45\,mm pitch, and is read out by a 64-channel VATA450.3 ASIC.  This instrument would nominally fit within a 4U volume and have a power draw of only 5\,W, allowing deployment in a low-inclination LEO on a 6U CubeSat bus.  Due largely to the low spacecraft mass and high efficiency of the compact CZT Compton telescope, initial simulations indicate a sensitivity somewhat better than the much larger INTEGRAL and CGRO instruments, with increased field of view and comparable angular resolution.

\section{Conclusions}

CubeSats have now proven themselves to be valuable additions to space-based gamma-ray astronomy.  In the future, 
6U, 12U, 16U, and 27U CubeSat standards will become more commonplace and open up even more science capabilities by allowing for larger and more sophisticated instruments. Standardized buses that can be commercially procured are now becoming available from companies such as NanoAvionics, Blue Canyon Technologies\footnote{\url{https://bluecanyontech.com}}, Terran Orbital\footnote{\url{https://https://terranorbital.com}}, Space Dynamics Laboratory\footnote{\url{https://www.sdl.usu.edu}}, and many others. The avionics packages for these busses, typically 1U--2U in size, have accumulated significant space flight heritage, and leave ample mass and volume for the science payload. 

Improvements in ground station infrastructures and sharing of resources such as S-band antennas, as foreseen by the HERMES team, will allow expanded data downlink capabilities. Advances in on-board processing, such as Edge AI  computing, will permit more sophisticated on-board data analysis for increasingly complex payloads, also reducing telemetry requirements.  

It will be essential that CubeSat mission data are archived and widely accessible, as planned by the BurstCube team. Tools and infrastructures being developed for rapid event notification `brokering' should be flexible enough to accommodate trigger information e.g. from CubeSat hardware beacons. 

The modularity, scalability and increasing performance of CubeSats will inevitably drive new science use cases that cannot be performed with traditional large spacecraft. The contributions of CubeSats to gamma--ray astronomy are already significant, and the future will undoubtedly be full of many more exciting and innovative missions.

\newpage

%\section{Cross-References \textit{(if applicable)}}
%Include a list of related entries from the handbook here that may be of further interest to the readers.

\section{Acknowledgements}
PB acknowledges support from the Laboratory Directed Research and Development program of Los Alamos National Laboratory under project number 20210047DR.  DM acknowledges support from Science Foundation Ireland grant 19/FFP/6777.

\end{document}